\documentclass[apl,amsmath,amssymb,amnsfonts,twocolumn,a4paper]{revtex4-1}
\usepackage{etex}
\usepackage[T1]{fontenc}
\usepackage{geometry}
\usepackage{graphicx}
\usepackage{color, colortbl}
\usepackage[usenames,dvipsnames]{pstricks}
\usepackage{epsfig}

\usepackage{setspace}
\usepackage{booktabs}
\usepackage[strict]{changepage}

\begin{document}

\title{Influence of grain size and exchange interaction on the LLB modeling procedure} 

\author{Christoph Vogler}
\email{christoph.vogler@tuwien.ac.at}
\affiliation{Institute of Solid State Physics, TU Wien, Wiedner Hauptstrasse 8-10, 1040 Vienna, Austria}
\affiliation{Institute of Analysis and Scientific Computing, TU Wien, Wiedner Hauptstrasse 8-10, 1040 Vienna, Austria}

\author{Claas Abert}
\author{Florian Bruckner}
\author{Dieter Suess}
\affiliation{Christian Doppler Laboratory for Advanced Magnetic Sensing and Materials, Institute for Solid State Physics, TU Wien, Wiedner Hauptstrasse 8-10, 1040 Vienna, Austria}

\author{Dirk Praetorius}
\affiliation{Institute of Analysis and Scientific Computing, TU Wien, Wiedner Hauptstrasse 8-10, 1040 Vienna, Austria}

\date{\today}

\begin{abstract}
  Reliably predicting bit-error rates in realistic heat-assisted magnetic recording simulations is a challenging task. Integrating the Landau-Lifshitz-Bloch (LLB) equation can reduce the computational effort to determine the magnetization dynamics in the vicinity of the Curie temperature. If one aims that these dynamics coincide with trajectories calculated from the atomistic Landau-Lifshitz-Gilbert equation, one has to carefully model required temperature dependent material functions such as the zero-field equilibrium magnetization as well as the parallel and normal susceptibilities. We present an extensive study on how these functions depend on grain size and exchange interactions. We show that, if the size or the exchange constant of a reference grain is modified, the material functions can be scaled, according to the changed Curie temperature, yielding negligible errors. This is shown to be valid for volume changes of up to $\pm 40$\,\% and variations of the exchange constant of up to $\pm 10$\,\%. Besides the temperature dependent material curves, computed switching probabilities also agree well with probabilities separately determined for each system. Our study suggest that there is no need to recalculate the required LLB input functions for each particle. Within the presented limits it is sufficient to scale them to the Curie temperature of the altered system.
\end{abstract}

\maketitle 

\section{Introduction}
\label{sec:intro}
With the increasing importance of heat-assisted magnetic recording (HAMR), high temperature micromagnetics have become an essential topic. Solving the Landau-Lifshitz-Gilbert (LLG) equation within a finite element framework cannot satisfy the demands, which arise at fast varying temperatures near the Curie point $T_{\mathrm{C}}$, because the magnitude of the magnetization is kept constant. At a fixed temperature, below $T_{\mathrm{C}}$, one could in principle use material parameters, which are adjusted to the specific simulation temperature, in order to compute the correct magnetization dynamics. Once the temperature starts to vary the LLG fails, due to the lack of longitudinal magnetization relaxation. In such a case one must use an atomistic discretization of the magnetic particle. Then, the phase transition from the ferromagnetic to the paramagnetic state at $T_{\mathrm{C}}$ follows from averaging over the spin ensemble. This procedure is computationally expensive, and thus as an alternative strategy one can solve the Landau-Lifshitz Bloch (LLB) equation \cite{garanin_fokker-planck_1997,garanin_thermal_2004,evans_stochastic_2012}. The LLB needs temperature dependent material functions, like the zero field equilibrium magnetization $m_{\mathrm{e}}$ and the longitudinal and perpendicular susceptibilities $\widetilde{\chi}_\parallel$ and $\widetilde{\chi}_\perp$ as an input. But after having obtained all requirements the LLB can be solved in a single spin approach without any mesh, which is computationally cheap \cite{chubykalo-fesenko_dynamic_2006,bunce_laser-induced_2010,volger_llb}.

To correctly model all finite size effects the temperature dependent material functions must be determined for each system size or composition. Especially, for HAMR simulations this is a crucial restriction, if one aims to consider size or $T_{\mathrm{C}}$ distributions of the recording grains. In this work we intend to investigate in which limits material functions, which were computed or measured for a specific system, can be reused for other systems, by comparing atomistic LLG and LLB simulation results. In detail, we analyze the effect of the system size and the exchange constant. We hope this study to become a LLB modeling guideline, which helps to estimate the error that occurs if one reuses temperature dependent material functions. Further, it should help to minimize these errors with little effort.

\section{Model}
\label{sec:model}
The LLB equation was designed to consider the longitudinal relaxation of the magnetization in a magnetic particle, without the need for an atomistic discretization. Many publications confirm its validity~\cite{garanin_thermal_2004,chubykalo-fesenko_dynamic_2006,atxitia_micromagnetic_2007,kazantseva_towards_2008,chubykalo-fesenko_dynamic_2006,schieback_temperature_2009,bunce_laser-induced_2010,evans_stochastic_2012,mcdaniel_application_2012,greaves_magnetization_2012,mendil_resolving_2014,volger_llb}. Our model uses the LLB, where the magnetization magnitude preserves the Boltzmann distribution up to the Curie temperature. It was formulated in Ref.~\cite{evans_stochastic_2012} per:
\begin{eqnarray}
\label{eq:LLB}
  \frac{d \boldsymbol{m}}{dt}= &-&\mu_0{\gamma'}\left( \boldsymbol{m}\times \boldsymbol{H}_{\mathrm{eff}}\right) \nonumber \\
  &-&\frac{\alpha_\perp\mu_0 {\gamma'}}{m^2} \left \{ \boldsymbol{m}\times \left [ \boldsymbol{m}\times \left (\boldsymbol{H}_{\mathrm{eff}}+\boldsymbol{\xi}_{\perp}  \right ) \right ] \right \}\nonumber \\
  &+&\frac{\alpha_\parallel  \mu_0{\gamma'}}{m^2}\boldsymbol{m}\left (\boldsymbol{m}\cdot\boldsymbol{H}_{\mathrm{eff}}  \right )+\boldsymbol{\xi}_{\parallel},
\end{eqnarray}
where $\gamma'$ is the reduced electron gyromagnetic ratio ($\gamma'=|\gamma_{\mathrm{e}}|/(1+\lambda^2)$ with $|\gamma_{\mathrm{e}}|=1.76086\cdot10^{11}$\,(Ts)$^{-1}$), $\mu_0$ is the vacuum permeability and $\alpha_\parallel$ and $\alpha_\perp$ are the longitudinal and perpendicular dimensionless damping constants, respectively. With $M_0$ being the saturation magnetization at zero temperature, the reduced magnetization is $\boldsymbol{m}=\boldsymbol{M}/M_0$. Thermal fluctuations are considered with thermal fields $\boldsymbol{\xi}_{\parallel}$ and $\boldsymbol{\xi}_{\perp}$. The field components are white noise random numbers. The effective field $\boldsymbol{H}_{\mathrm{eff}}$ in Eq.~\ref{eq:LLB} contains the external field $\boldsymbol{H}_{\mathrm{ext}}$, the anisotropy field along the $z$ direction
\begin{equation}
  \label{eq:Hani}
   \boldsymbol{H}_\mathrm{ani}=\frac{1}{\widetilde{\chi}_{\perp}(T)}\left( m_x\boldsymbol{e}_{x}+m_y\boldsymbol{e}_{y}\right),
\end{equation}
and the internal exchange field 
\begin{equation}
\label{eq:blochField}
 \boldsymbol{H}_{\mathrm{J}}=\begin{cases} \frac{1}{2\widetilde{\chi}_{\parallel}(T)}\left( 1-\frac{m^2}{m^2_{\mathrm{e}}(T)} \right)\boldsymbol{m} & T\lesssim T_{\mathrm{C}}\\ -\frac{1}{\widetilde{\chi}_{\parallel}(T)} \left( 1+\frac{3}{5}\frac{T_{\mathrm{C}}}{T-T_{\mathrm{C}}}m^2 \right)\boldsymbol{m}& T\gtrsim T_{\mathrm{C}}.\end{cases}
\end{equation}
We represent each particle with one single magnetization vector in our study. Hence, the effective field does not contain an exchange field. In Eqs.~\ref{eq:Hani} and \ref{eq:blochField} the longitudinal and perpendicular susceptibilities $\widetilde{\chi}_{\parallel}$ and $\widetilde{\chi}_{\perp}$ and the zero field equilibrium magnetization $m_{\mathrm{e}}$ are temperature dependent material functions, which have to be precomputed, in order to obtain the correct dynamical high temperature behavior. As already mentioned, strictly speaking these functions are dependent on the system size and composition. We calculate $\widetilde{\chi}_{\parallel}(T)$, $\widetilde{\chi}_{\perp}(T)$ and $m_{\mathrm{e}}(T)$ from stochastic LLG simulations with an atomistic discretization by means of the code VAMPIRE~\cite{evans_atomistic_2014}. VAMPIRE solves for the time evolution of the spins $\boldsymbol{S}_k$ with constant magnitude per:
\begin{eqnarray}
 \label{eq:atomisticLLG}
 \frac{d\boldsymbol{S}_k}{dt}=&-& \gamma'\left \{\boldsymbol{S}_k \times\left ( \boldsymbol{H}_{\mathrm{eff},k}+\boldsymbol{\xi}_k \right )   \right \}\nonumber \\
 &-& \gamma' \lambda \left \{ \boldsymbol{S}_k \times \left [ \boldsymbol{S}_k \times \left (  \boldsymbol{H}_{\mathrm{eff},k}+\boldsymbol{\xi}_k\right ) \right ] \right \}.
\end{eqnarray}
Here, the effective field contains the external field, the anisotropy field and the exchange field. For more details about the models please refer to~\cite{volger_llb}. 
\section{finite size effects}
\label{sec:finite_size_effects}
\begin{figure}
\includegraphics{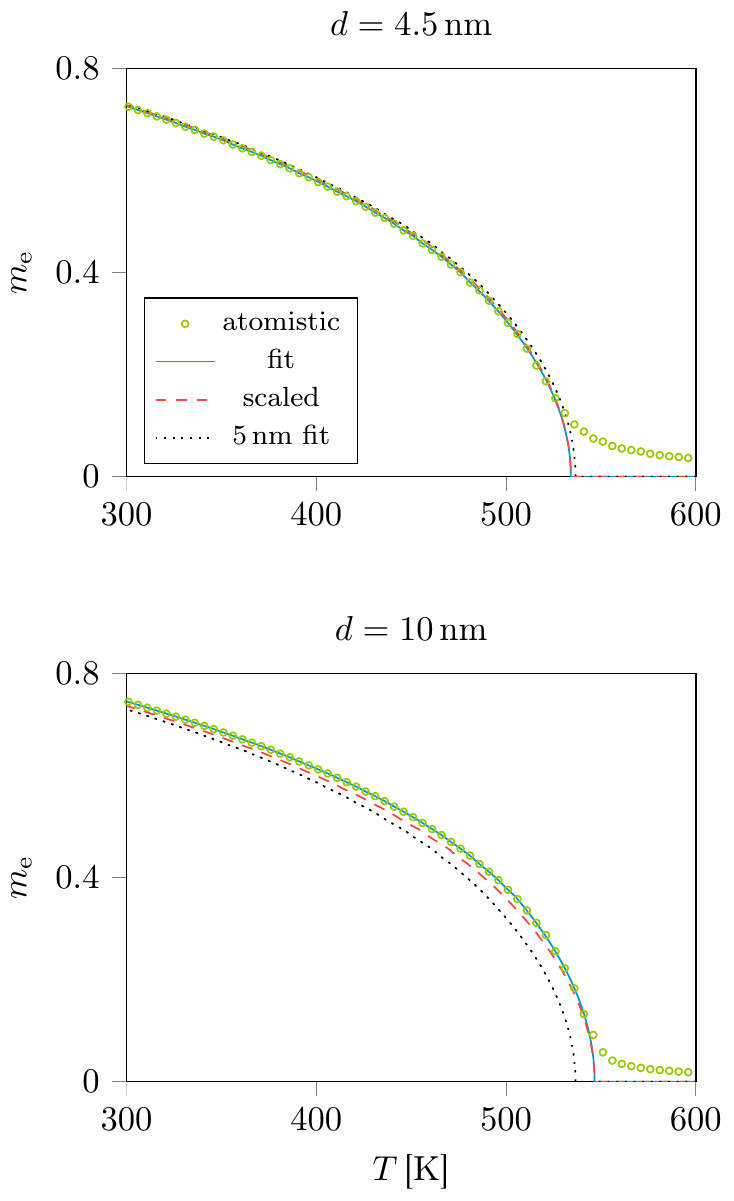}
  \caption{\small Zero field equilibrium magnetization $m_{\mathrm{e}}$ of a cylindrical particle with two different diameters and material parameters as given in Tab.~\ref{tab:mat}. Results of atomistic LLG simulations (green circles) and the corresponding infinite size fits (solid blue), as well as the $m_{\mathrm{e}}$ fit of the 5\,nm reference particle (dotted black) are plotted. The latter is scaled to the Curie temperature of the actual size (dashed red).}
  \label{fig:me_cylinder}	
\end{figure}

\begin{figure}
\includegraphics{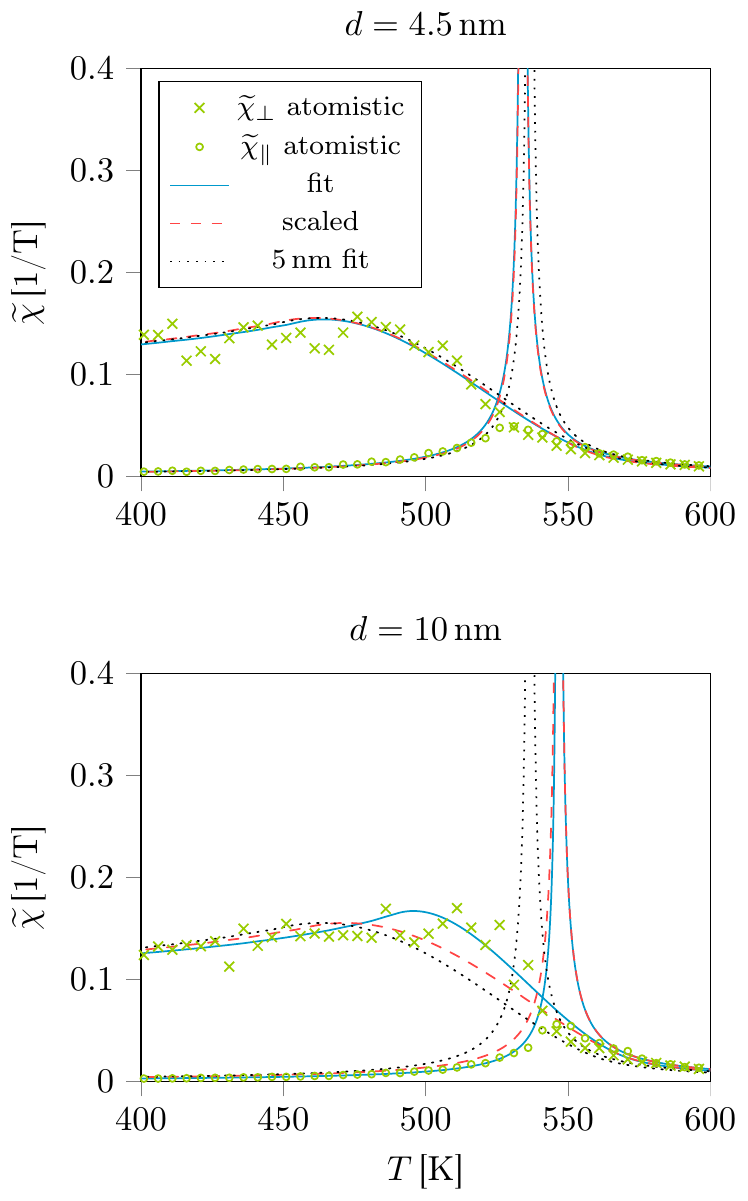}
  \caption{\small Longitudinal ($\widetilde{\chi}_{\parallel}$) and perpendicular ($\widetilde{\chi}_{\perp}$) susceptibilities of a cylindrical particle with two different diameters and material parameters given in Tab.~\ref{tab:mat}. Results of atomistic LLG simulations (green circles and crosses) and the corresponding infinite size fits (solid blue), as well as the susceptibility fits of the 5\,nm reference particle (dotted black) are plotted. The latter is scaled to the Curie temperature of the actual size (dashed red).}
  \label{fig:chi_cylinder}	
\end{figure}
We investigate how the functions $\widetilde{\chi}_{\parallel}(T)$, $\widetilde{\chi}_{\perp}(T)$ and $m_{\mathrm{e}}(T)$ depend on the diameter of a cylindrical particle with a constant height of 10\,nm. For each diameter, in a range of 3.5\,nm to 10\,nm, VAMPIRE simulations with a time step of 10$^{-15}$\,s are performed at various temperatures ($0-800$\,K). At each temperature 100 trajectories consisting of 20000 equilibration steps and 20000 simulation steps are computed in the absence of any external field. Averaging the magnetization components $m_\eta$ over all simulation steps yields the zero field equilibrium magnetization. Note, the magnetization components are calculated from the ensemble of $N$ spins in the particle per:
\begin{equation}
 m_\eta=\frac{1}{N}\sum_{i=k}^{N}S_{\eta,k}.
\end{equation}
From the fluctuations of these components one can compute $\widetilde{\chi}_{\parallel}(T)$ and $\widetilde{\chi}_{\perp}(T)$. Finally, the three temperature dependent functions are fitted. The detailed procedure, how to properly extract the fits from atomistic LLG simulations can be found in Refs.~\cite{kazantseva_towards_2008,volger_llb}. 

\begin{table}
  \centering
  \begin{tabular}{c c c c c}
    \toprule
    \toprule
      $K_1$\,[J/m$^3$] & $J_{\mathrm{S}}$\,[T] & $A_{\mathrm{ex}}$\,[pJ/m] & $a$\,[nm] & $\lambda$\\
    \midrule
      $6.6\times10^6$ & 1.43 & 21.58 & 0.24 & 0.1\\
    \bottomrule
    \bottomrule
  \end{tabular}
  \caption{\small Material properties of the reference grain. $K_1$ is the uniaxial anisotropy constant, $J_{\mathrm{S}}$ is the saturation polarization and $\lambda$ is the dimensionless damping constant. $A_{\mathrm{ex}}$ denotes the exchange constant and $a$ is the lattice constant in the atomistic model. All parameters are zero temperature values.}
  \label{tab:mat}
\end{table}

The choice of the size of the smallest particle was motivated by the findings of Ref.~\cite{ellis_switching_2015}, which suggest that for even smaller particles the LLB equation, which is actually derived in the bulk regime, is not valid any more.

In this section we want to focus on the differences originating from varying cylinder diameters, and thus particle volumes. More precisely, we define a reference particle with a cylinder diameter of 5\,nm and the material parameters of Tab.~\ref{tab:mat}. For other system system sizes the Curie temperature varies due to finite size effects. Hence, the temperature dependent material functions vary too. Since it is time consuming to extract the correct functions, reusing existing ones from the 5\,nm particle would be very helpful. As a consequence, we compare the directly fitted $\widetilde{\chi}_{\parallel}(T)$, $\widetilde{\chi}_{\perp}(T)$ and $m_{\mathrm{e}}(T)$ curves with the 5\,nm curves, after scaling (or shifting) them to the new Curie temperature. For example, to analyze the difference of $m_{\mathrm{e}}(T)$ of the 5\,nm system and the 10\,nm system, we directly calculate both fits from atomistic simulations. After that, we scale the 5\,nm equilibrium magnetization curve per:
\begin{equation}
\label{eq:scaling}
 m_{\mathrm{e,sc,10\,nm}}(T)=m_{\mathrm{e,at,5\,nm}}\left(T\frac{T_{\mathrm{C,10\,nm}}}{T_{\mathrm{C,5\,nm}}}\right)
\end{equation}
or shift it per:
\begin{equation}
\label{eq:shifting}
 m_{\mathrm{e,sh,10\,nm}}(T)=m_{\mathrm{e,at,5\,nm}}\left(T+\Delta T_{\mathrm{C}}\right ).
\end{equation}
Here, ``at'' indicates the atomistic fit, ``sc'' indicates the scaled fit and ``sh'' the shifted fit. Figure~\ref{fig:me_cylinder} exemplarily illustrates one system where the scaled magnetization agrees very well with the atomistic data and one where deviations are observable. The same comparison is shown for the susceptibilities in Fig.\ref{fig:chi_cylinder}. 

To quantify the agreement, we compute the mean squared displacement (MSD) which is defined as:
\begin{equation}
 \left \langle \left(a-b\right)^2 \right\rangle = \frac{1}{N}\sum_{i=1}^{N} \left[a_i-b_i \right]^ 2.
\end{equation}
The sum is performed over all $N$ data points in a temperature range from 300\,K to 800\,K ($\Delta T=5$\,K), which is relevant to HAMR. In particular, we are interested in the following MSD ratios:
\begin{itemize}
 \item $\mathrm{rMSD}_{\mathrm{sc}}(x)=\frac{\left \langle \left(x-x_{\mathrm{sc}}\right)^2 \right\rangle}{\left \langle \left(x-x_{\mathrm{at}}\right)^2 \right\rangle}$: ratio of the MSD of atomistic data $x$ and the scaled 5\,nm fit $x_{\mathrm{sc}}$ and the MSD of atomistic data $x$ and the corresponding fit $x_{\mathrm{at}}$ for the specific size.
 \item $\mathrm{rMSD}_{\mathrm{sh}}(x)=\frac{\left \langle \left(x-x_{\mathrm{sh}}\right)^2 \right\rangle}{\left \langle \left(x-x_{\mathrm{at}}\right)^2 \right\rangle}$: ratio of the MSD of atomistic data $x$ and the shifted 5\,nm fit $x_{\mathrm{sh}}$ and the MSD of atomistic data $x$ and the corresponding fit $x_{\mathrm{at}}$ for the specific size.
\end{itemize}
$x$ is a placeholder for $\widetilde{\chi}_{\parallel}(T)$, $\widetilde{\chi}_{\perp}(T)$ or $m_{\mathrm{e}}(T)$, respectively. The MSD ratios represent the quality of the scaling and shifting approach. Low ratios indicate that the error is small if materials curves are scaled or shifted instead of directly fitted.

In the case of $m_{\mathrm{e}}(T)$ the MSD is truncated at $T_{\mathrm{C}}$, because per definition the equilibrium magnetization fits are zero above. Another special case appears for $\widetilde{\chi}_{\parallel}(T)$, which diverges at $T_{\mathrm{C}}$. Hence, a temperature range from $T_{\mathrm{C}}-10$\,K to $T_{\mathrm{C}}+10$\,K is excluded in the MSD calculation. 

\begin{figure}
\includegraphics{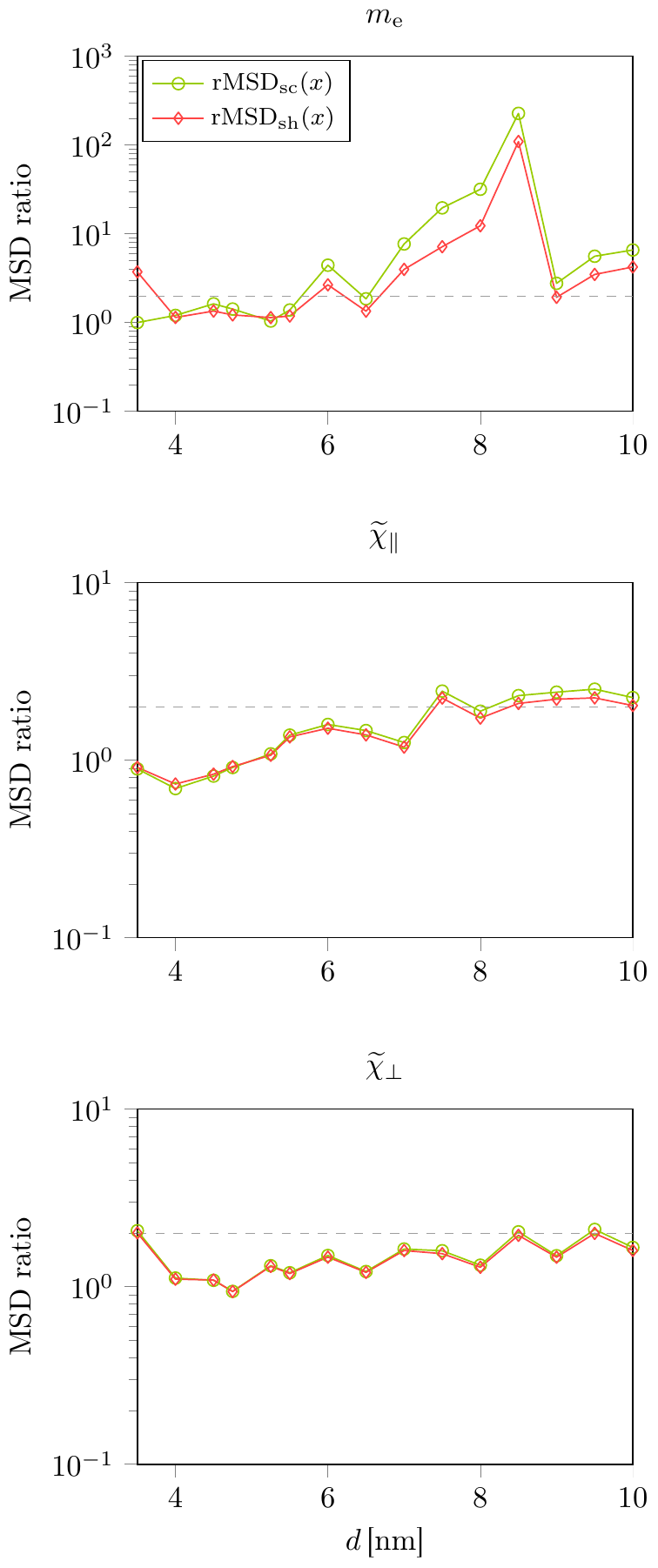}
  \caption{\small Mean squared displacement (MSD) ratios of the scaled ($\mathrm{rMSD}_{\mathrm{sc}}(x)$) and shifted ($\mathrm{rMSD}_{\mathrm{sh}}(x)$) temperature dependent material functions for various particle diameters. Here, $x$ is a placeholder for $\widetilde{\chi}_{\parallel}(T)$, $\widetilde{\chi}_{\perp}(T)$ and $m_{\mathrm{e}}(T)$, respectively.}
  \label{fig:cylinder_meanSquaredDisp}	
\end{figure}

Figure~\ref{fig:cylinder_meanSquaredDisp} displays the MSD ratios of the three temperature dependent functions for all investigated cylinder diameters. In the case of the equilibrium magnetization it can be seen that from 3.5\,nm to 7\,nm diameter the MSD ratios of for the scaled and the shifted $m_{\mathrm{e}}(T)$ fit are within one magnitude. In a smaller range from 4\,nm to 5.5\,nm the MSD ratios are even below 2.0. Having in mind that one cannot distinguish the direct and the scaled fit in Fig.~\ref{fig:me_cylinder}a the error of the scaled and shifted equilibrium magnetizations seems to be negligible. rMSD$_{\mathrm{sc}}(\widetilde{\chi}_{\parallel})$ and rMSD$_{\mathrm{sh}}(\widetilde{\chi}_{\parallel})$ show a small error up to a diameter of 7.5\,nm. The MSD ratios are below 2.0 for all analyzed particle sizes in the case of the transversal susceptibility. The reason is, that $\widetilde{\chi}_{\perp}$ is rather noisy, as Fig~\ref{fig:chi_cylinder} points out. It has to be noted that both the scaling and the shifting of the 5\,nm functions yield small errors within the examined temperature range off 300\,K to 800\,K. For lower temperatures rMSD$_{\mathrm{sh}}(m_{\mathrm{e}})$ would become larger, because due to the shifting according to Eq.~\ref{eq:shifting} the reduced equilibrium magnetization at 0\,K would not be one. But low temperatures are of little interest for HAMR.

\subsection{switching probability}
\begin{figure}
\includegraphics{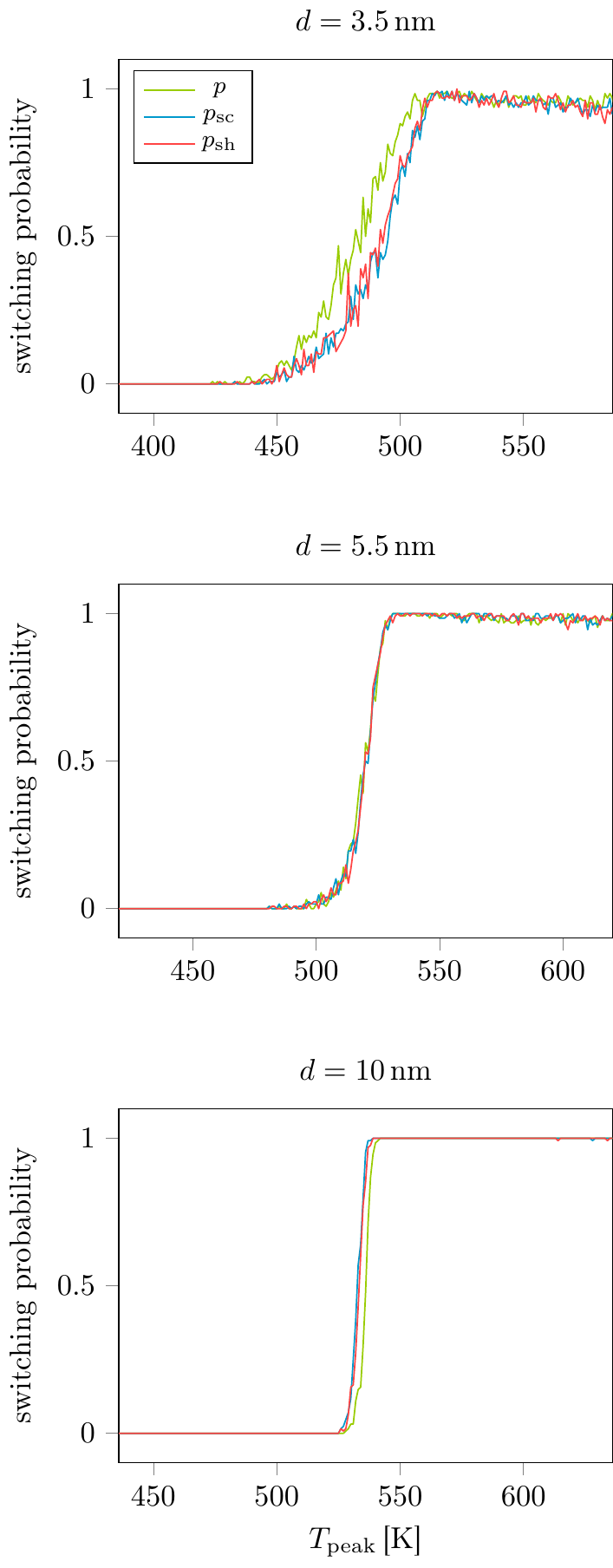}
  \caption{\small Switching probability versus peak temperature curves for various particle diameters. Each plot compares LLB simulation results with $\widetilde{\chi}_{\parallel}(T)$, $\widetilde{\chi}_{\perp}(T)$ or $m_{\mathrm{e}}(T)$ input functions, obtained separately for each grain size ($p$), with probabilities computed from the scaling ($p_{\mathrm{sc}}$) and shifting ($p_{\mathrm{sh}}$) approach, respectively.}
  \label{fig:cylinder_prob}	
\end{figure}
The main goal of HAMR simulations is to efficiently calculate switching probabilities and bit error rates. Hence, we test if the scaled and shifted $\widetilde{\chi}_{\parallel}(T)$, $\widetilde{\chi}_{\perp}(T)$ or $m_{\mathrm{e}}(T)$ functions yield the same switching behavior in LLB simulations as separately calculated material curves. A Gaussian shaped heat pulse is applied to the grains per:
\begin{equation}
\label{eq:gauss_profile_PLSR}
 T(t)=T_{\mathrm{min}}+\left ( T_{\mathrm{peak}}-T_{\mathrm{min}} \right ) e^{-\frac{\left (t-t_0 \right )^2}{\tau^2}},
\end{equation}
with $T_{\mathrm{min}}=270$\,K and $\tau=200$\,ps. Additionally, a constant external magnetic field with 0.8\,T assists the switching of the particle from its original state, with the magnetization pointing in $z$ direction, to the $-z$ direction. At each peak temperature 128 switching trajectories are simulated, by means of Eq.~\ref{eq:LLB}. Afterwards the ratio of switched and not switched particles is evaluated, yielding the switching probability. For various particle sizes the simulations are performed with the original temperature dependent material curves, the shifted and the scaled functions. Figure~\ref{fig:cylinder_prob} exemplarily illustrates the results for three particle sizes. The smallest and the largest investigated grains clearly show significant deviations between the switching probabilities $p$, computed with the directly fitted material functions for the appropriate size, and the probabilities of scaled and shifted functions $p_{\mathrm{sc}}$ and $p_{\mathrm{sh}}$. Although, the switching probabilities at high peak temperatures agree well, the transition cannot be reproduced. In the case of a 5.5\,nm particle diameter all probability curves coincide. Note, to facilitate comparison, the $x$ axes in Fig.~\ref{fig:cylinder_prob} have different ranges. The intention was to center the probability transitions.

\begin{figure}
\includegraphics{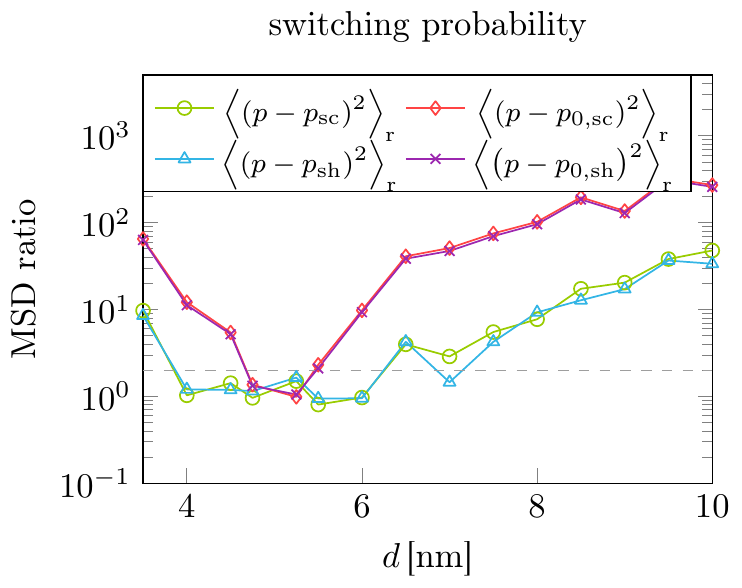}
  \caption{\small MSD ratios of the switching probability for various grain diameters. $p$ denotes switching probabilities obtained from LLB simulations with separately fitted material functions for each size, $p_{\mathrm{sc}}$ and $p_{\mathrm{sh}}$ represent LLB simulation results with scaled and shifted $\widetilde{\chi}_{\parallel}(T)$, $\widetilde{\chi}_{\perp}(T)$ or $m_{\mathrm{e}}(T)$ curves (5\,nm reference grain). $p_{\mathrm{0,sc}}$ and $p_{\mathrm{0,sh}}$ indicate directly scaled and shifted probability curves of the 5\,nm grain.}
  \label{fig:prob_cylinder_meanSquaredDisp}	
\end{figure}

To quantify the results we compute the MSD of the switching probabilities obtained from direct and scaled as well as direct and shifted material curves $\left \langle \left (p-p_{\mathrm{sc}}\right)^2 \right\rangle$ and $\left \langle \left (p-p_{\mathrm{sh}}\right)^2 \right\rangle$, respectively. More precisely, the MSD ratio of these quantities and $\left \langle \left (p_i-p_j\right)^2 \right\rangle$ are evaluated. Since the probabilities have a stochastic nature the repeated simulation with the same input parameters yields slightly different results. Hence, the MSD of the repeated computation of $p$ is the basis of our analysis, because it is assumed to be the smallest possible. Figure~\ref{fig:prob_cylinder_meanSquaredDisp} points out that the MSD ratios of all grain sizes until 8\,nm diameter are within one magnitude, for both LLB simulations with the scaled and the shifted material functions. In a wide range, from 4\,nm to 6\,nm, the ration is clearly below 2.0, which is an excellent agreement. Notably, the scaling and shifting approaches yield the correct dynamical behavior for volume changes of up to about $\pm 40$\,\%. This also coincides well with the findings of Fig.~\ref{fig:cylinder_meanSquaredDisp}.

Instead of scaling or shifting $\widetilde{\chi}_{\parallel}(T)$, $\widetilde{\chi}_{\perp}(T)$ and $m_{\mathrm{e}}(T)$ and calculating switching probabilities we could directly scale or shift the switching probability curve of the 5\,nm particle corresponding to the modified $T_{\mathrm{C}}$ value of other system sizes (equivalently to Eqs.~\ref{eq:scaling} and \ref{eq:shifting}). This procedure is computationally very cheap, but it can, of course, not capture the finite size effects of large size variations. Nevertheless, Fig.~\ref{fig:prob_cylinder_meanSquaredDisp} reveals that for minor changes of the cylinder diameter of $\pm 0.25$\,nm the MSD ratios $\left \langle \left (p-p_{\mathrm{0,sc}}\right)^2 \right\rangle_{\mathrm{r}}$ and $\left \langle \left (p-p_{\mathrm{0,sh}}\right)^2 \right\rangle_{\mathrm{r}}$ are as low as for the recalculated probability curves. Remarkably, this corresponds to a volume change of $\pm 10$\,\%. 

\subsection{modeling strategy}
\begin{figure}
\includegraphics{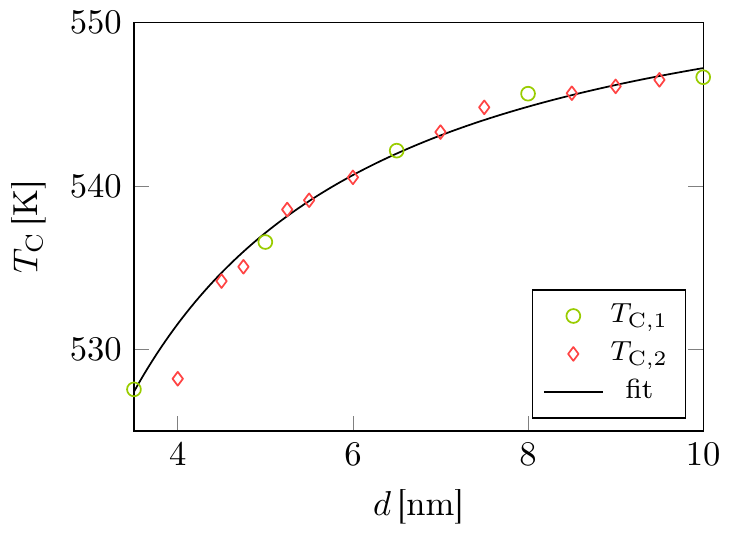}
  \caption{\small Finite size Curie temperature $T_{\mathrm{C}}(d)$ for various particles sizes, obtained from atomistic LLG simulations. The simulated $T_{\mathrm{C,1}}(d)$ are fitted with the finite size scaling law (Eq.~\ref{eq:finite_size_scaling_law}) with $T_{\mathrm{C}}^\infty$, $\Lambda$ and $d_0$ being fit parameters. The fit agrees well with various finite size Curie temperatures $T_{\mathrm{C,2}}(d)$, which are not used for the fit.}
  \label{fig:fit_Tc}	
\end{figure}
In the above section it was shown that the temperature dependent material functions $\widetilde{\chi}_{\parallel}(T)$, $\widetilde{\chi}_{\perp}(T)$ and $m_{\mathrm{e}}(T)$ of one specific grain, which are required to integrate the LLB equation (Eq.~\ref{eq:LLB}), are sufficient to predict the dynamical behavior of particles with similar sizes. To make use of the demonstrated scaling or shifting approach one must know the Curie temperatures of the involved systems. According to Ref.~\cite{hovorka_curie_2012} $T_{\mathrm{C}}(d)$ follows the finite size scaling law:
\begin{equation}
\label{eq:finite_size_scaling_law}
 \frac{T_{\mathrm{C}}^{\infty}-T_{\mathrm{C}}(d)}{T_{\mathrm{C}}^{\infty}}=\left( \frac{d_0}{d} \right )^\Lambda,
\end{equation}
where $T_{\mathrm{C}}^{\infty}$ is the bulk Curie temperature and $\Lambda$ and $d_0$ are material and model dependent quantities, respectively. $T_{\mathrm{C}}^{\infty}$ and $\Lambda$ could in principle be determined from the finite size scaling analysis \cite{binder_applications_1997,hovorka_curie_2012}, but we suggest to use them, together with $d_0$, as fit parameters. As Fig.~\ref{fig:fit_Tc} indicates, we propose to compute $T_{\mathrm{C}}(d)$ for a few grain sizes from atomistic LLG simulations ($T_{\mathrm{C,1}}(d)$ in Fig.~\ref{fig:fit_Tc}). Afterwards these data can be fitted with Eq.~\ref{eq:finite_size_scaling_law} and the Curie point of other particle sizes can be estimated from the fit function (see Fig.~\ref{fig:fit_Tc}). With the known value of $T_{\mathrm{C}}(d)$ the scaling or shifting approach of the previous section can be easily applied. This strategy allows to efficiently and accurately model arbitrary grain sizes with the LLB equation, within the presented limitations.

Additionally to the cylindrical particle we investigated a cube with various edge lengths (again 3.5\,nm to 10\,nm) and performed all so far shown calculations. The results are not explicitly given, because based on the volume to surface ratio the cuboid particle revealed the same scaling behavior as the cylindrical grain.

\section{exchange interaction effects}
\label{sec:exchange_interaction_effects}
\begin{figure}
\includegraphics{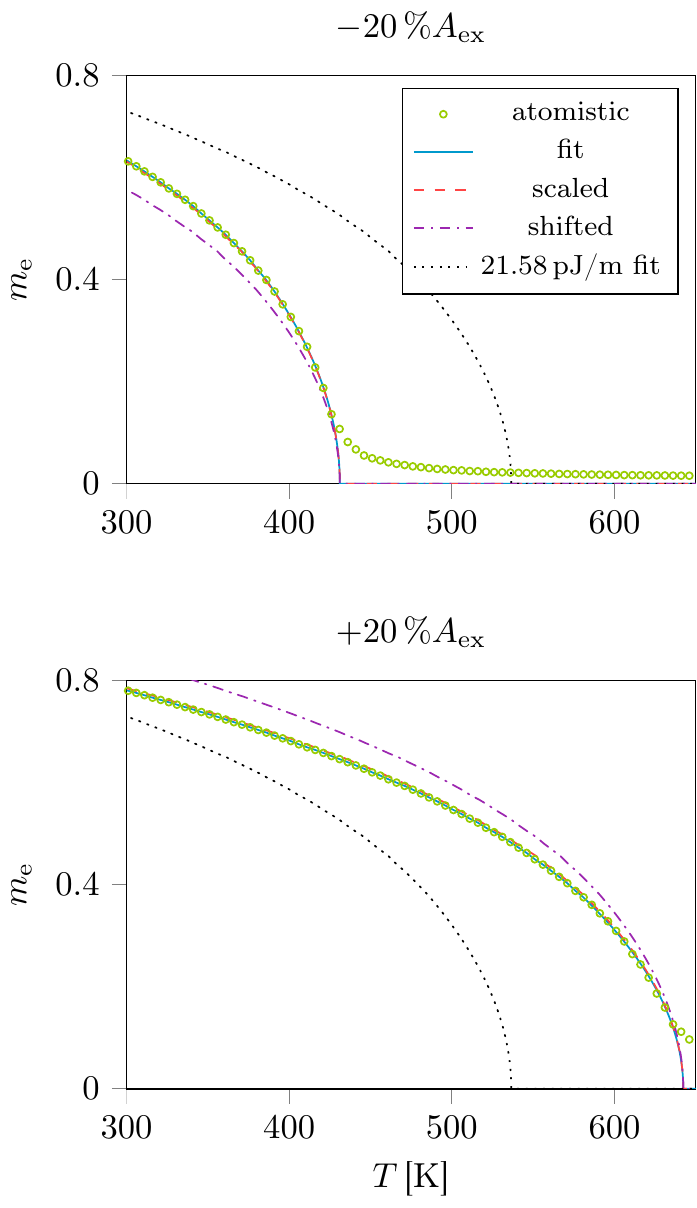}
  \caption{\small Zero field equilibrium magnetization $m_{\mathrm{e}}$ of a cylindrical particle with 5\,nm diameter and two different exchange constants, based on the material parameters of Tab.~\ref{tab:mat}. Results of atomistic LLG simulations (green circles) and the corresponding infinite size fits (solid blue), as well as the $m_{\mathrm{e}}$ fit of the reference particle with $A_{\mathrm{ex}}=21.58$\,pJ/m (dotted black) are plotted. The latter is scaled and shifted to the Curie temperature of the actual changed exchange constant (dashed red and chain dotted pink), respectively.}
  \label{fig:me_AiexVariations}	
\end{figure}

\begin{figure}
\includegraphics{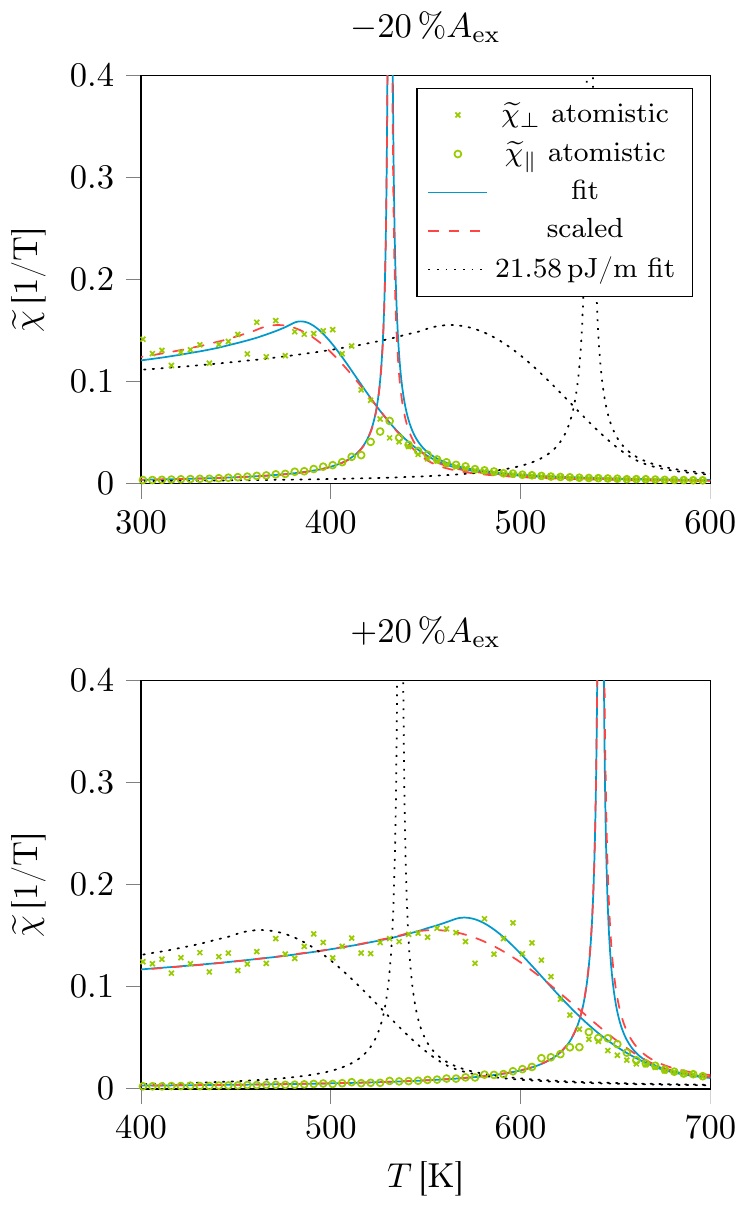}
  \caption{\small Longitudinal ($\widetilde{\chi}_{\parallel}$) and perpendicular ($\widetilde{\chi}_{\perp}$) susceptibilities of a 5\,nm cylindrical particle with two different exchange constants based on the the material parameters of Tab.~\ref{tab:mat}. Results of atomistic LLG simulations (green circles and crosses) and the corresponding infinite size fits (solid blue), as well as the susceptibility fits of the reference particle with $A_{\mathrm{ex}}=21.58$\,pJ/m (dotted black) are plotted. The latter is scaled to the Curie temperature of the new exchange constant (dashed red).}
  \label{fig:chi_AiexVariations}	
\end{figure}

Size variations just slightly change the particle's Curie temperature, as for example shown in Fig.~\ref{fig:fit_Tc}. In order to reliably estimate bit error rates and areal storage densities in HAMR simulations $T_{\mathrm{C}}$ distributions must be considered. The main source of these distributions is a variation of the exchange interaction between the neighboring spins in a recording grain. In this section we investigate if the scaling or shifting strategy also works for changes of the exchange constant $A_{\mathrm{ex}}$. For this purpose, we analyze how the temperature dependent material functions of a cylindrical particle with a diameter of 5\,nm and a height of 10\,nm depend on $A_{\mathrm{ex}}$. As reference an exchange constant of $A_{\mathrm{ex}}=21.58$\,pJ/m is used, which is varied by up to $\pm 20$\,\%. Similar to Sec.~\ref{sec:finite_size_effects} we compare fits of $\widetilde{\chi}_{\parallel}(T)$, $\widetilde{\chi}_{\perp}(T)$ and $m_{\mathrm{e}}(T)$, obtained from atomistic LLG simulations, with scaled and shifted curves of the system with $A_{\mathrm{ex}}=21.58$\,pJ/m. The latter two are computed equivalently to Eqs.~\ref{eq:scaling} and \ref{eq:shifting} for exchange constants instead of particle diameters. 

\begin{figure}
\includegraphics{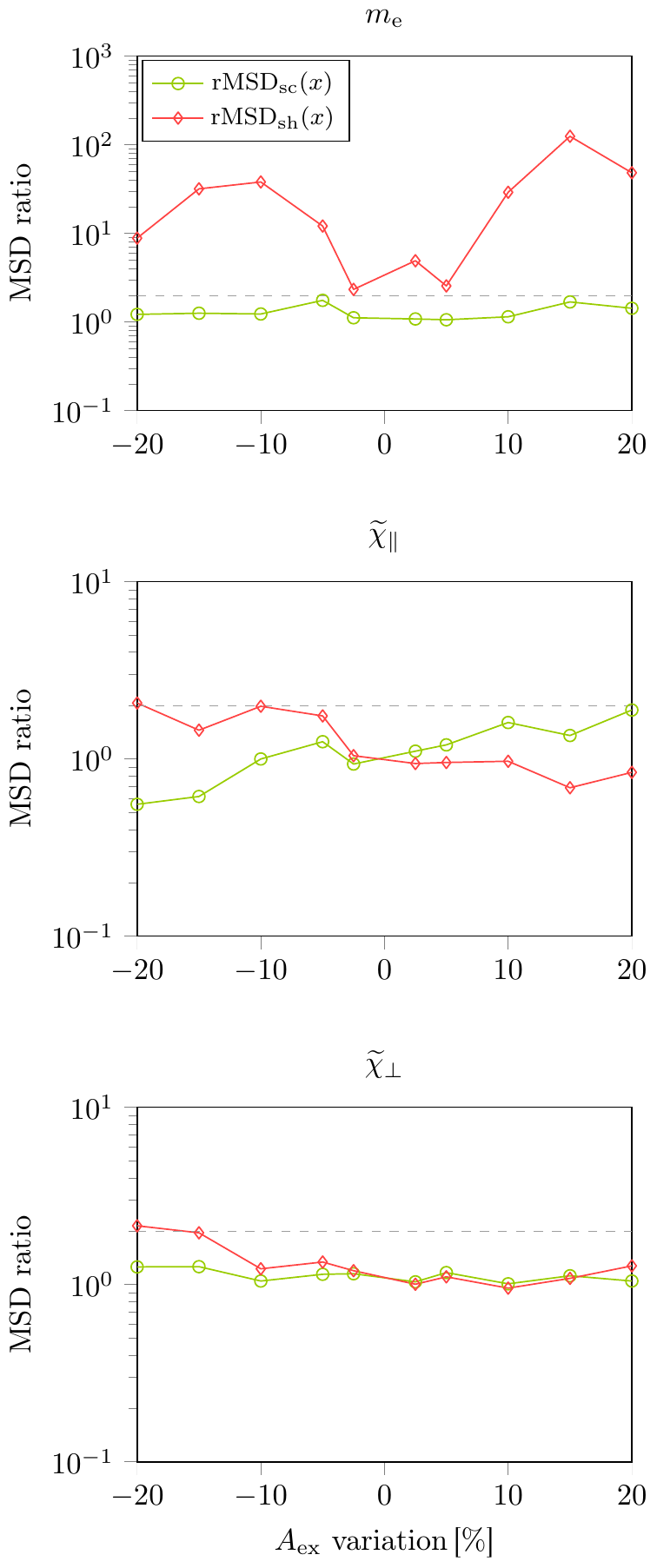}
  \caption{\small MSD ratios of the scaled ($\mathrm{rMSD}_{\mathrm{sc}}(x)$) and shifted ($\mathrm{rMSD}_{\mathrm{sh}}(x)$) temperature dependent material functions for various exchange constants. Here, $x$ is a placeholder for $\widetilde{\chi}_{\parallel}(T)$, $\widetilde{\chi}_{\perp}(T)$ and $m_{\mathrm{e}}(T)$, respectively.}
  \label{fig:cylinder_meanSquaredDisp_AiexVar}	
\end{figure}

In the case of the smallest and the largest analyzed exchange constants Figs.~\ref{fig:me_AiexVariations} and \ref{fig:chi_AiexVariations} exemplarily illustrate the equilibrium magnetization and the longitudinal and perpendicular susceptibilities, respectively. Despite the significant change of the Curie temperature, the scaled material curves agree surprisingly well with the atomistic data. In contrast to Sec.~\ref{sec:finite_size_effects}, the shifted $m_{\mathrm{e}}(T)$ curves show significant discrepancies. Due to the large shift of the Curie temperature the correct slope cannot be reproduced, as Fig.~\ref{fig:me_AiexVariations} points out. The ratios of the MSD of the atomistic data and the scaled or shifted $A_{\mathrm{ex}}=21.58$\,pJ/m fits confirm this trend, as displayed in Fig.~\ref{fig:cylinder_meanSquaredDisp_AiexVar}. The scaled material functions are almost identical to the atomistic results in the whole range of exchange constants. The MSD ratios of the shifted susceptibilities show the same agreement, but $\mathrm{rMSD}_{\mathrm{sh}}(m_{\mathrm{e}})$ is just within one magnitude for small deviations of the exchange constant. 

Nevertheless, the main finding is that the temperature dependent material functions of the scaling approach are as accurate as direct fits of atomistic data within the whole investigated range of $A_{\mathrm{ex}}$ values.

\subsection{switching probability}
\begin{figure}
\includegraphics{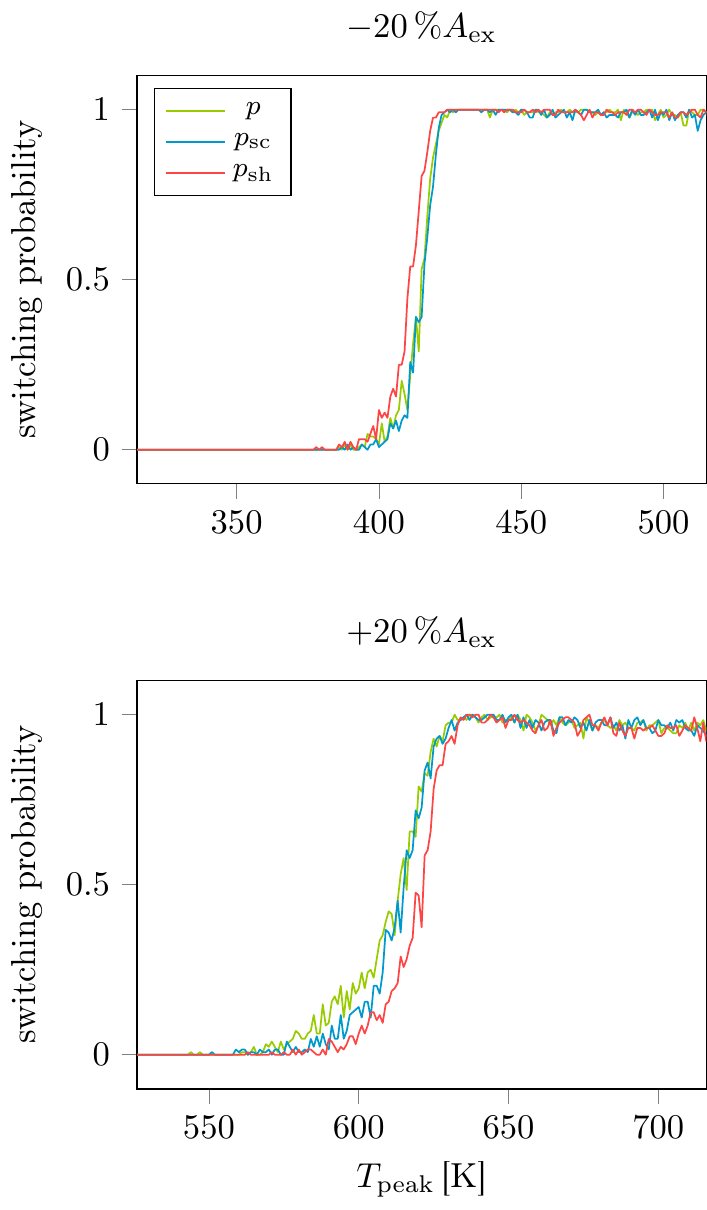}
  \caption{\small Switching probability versus peak temperature curves for various exchange constants. Each plot compares LLB simulation results with $\widetilde{\chi}_{\parallel}(T)$, $\widetilde{\chi}_{\perp}(T)$ or $m_{\mathrm{e}}(T)$ input functions, obtained separately for each exchange constant ($p$), with probabilities computed from the scaling ($p_{\mathrm{sc}}$) and shifting ($p_{\mathrm{sh}}$) approach, respectively.}
  \label{fig:prob_AiexVariations}	
\end{figure}

\begin{figure}
\includegraphics{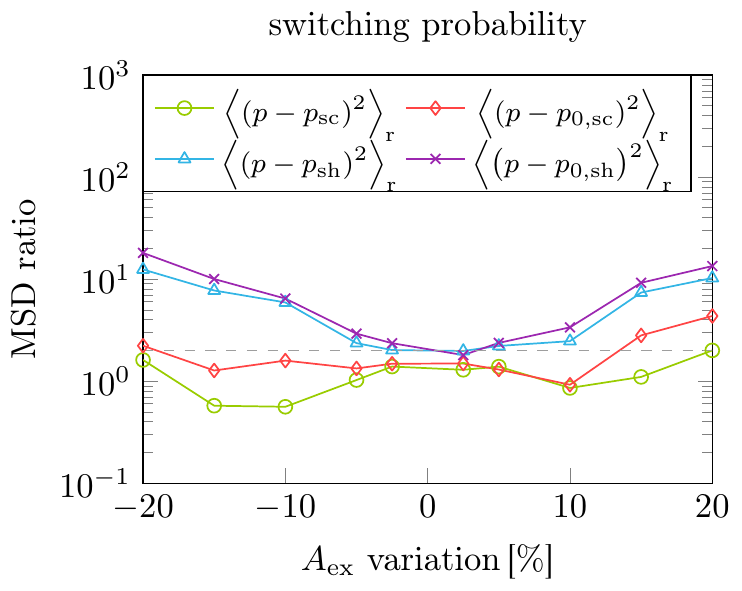}
  \caption{\small MSD ratios of the switching probability for various exchange constants. The same plots are shown as in Fig.~\ref{fig:prob_cylinder_meanSquaredDisp}. Here, the scaling and shifting is performed with respect to the exchange constants of the particles. The reference system is a cylindrical grain with 5\,nm diameter and 10\,nm height and an exchange constant of $A_{\mathrm{ex}}=21.58$\,pJ/m.
 }
  \label{fig:prob_cylinder_meanSquaredDisp_AiexVariations}	
\end{figure}

To confirm the good accordance of the scaling approach switching probabilities of the grains, as described in Sec.~\ref{sec:finite_size_effects}, are computed. The resulting probabilities for $A_{\mathrm{ex}}\pm 20$\,\% are shown in Fig.~\ref{fig:prob_AiexVariations}. In both cases the switching probability obtained from LLB simulations with the scaled material functions agrees better with $p$ of the directly fitted functions, than $p_{\mathrm{sh}}$. The agreement is worse for $+20\,\% A_{\mathrm{ex}}$ than for $-20\,\% A_{\mathrm{ex}}$. Figure~\ref{fig:prob_cylinder_meanSquaredDisp_AiexVariations} compares the MSD ratios of the switching probabilities in the whole range of $A_{\mathrm{ex}}$ variations. As expected, the scaling approach performs much better than the shifting approach. All MSD ratios $\left \langle \left (p-p_{\mathrm{sc}}\right)^2 \right\rangle_\mathrm{r}$ are below 2.0, with the exception of $+20\,\% A_{\mathrm{ex}}$. In contrast $\left \langle \left (p-p_{\mathrm{sh}}\right)^2 \right\rangle_\mathrm{r}$ is just comparable for $A_{\mathrm{ex}}$ variations up to $\pm 2.5$\,\%.

Another important finding is the fact, that scaling of the switching probability curve of a particle with $A_{\mathrm{ex}}=21.58$\,pJ/m, corresponding to the new Curie temperature yields an excellent MSD ratio (see $\left \langle \left (p-p_{\mathrm{0,sh}}\right)^2 \right\rangle_\mathrm{r}$ in Fig.~\ref{fig:prob_cylinder_meanSquaredDisp_AiexVariations}). This means, one has to calculate just the switching probabilities of a desired material and one can consider a change of the exchange constant, and thus $T_{\mathrm{C}}$, by scaling the probability curve per:
\begin{equation}
 \label{eq:scale_prob}
 \tilde{p}\left(T,T_{\mathrm{C}}\pm\Delta T_{\mathrm{C}}\right)=p\left( T\frac{T_{\mathrm{C}}\pm\Delta T_{\mathrm{C}}}{T_{\mathrm{C}}},T_{\mathrm{C}} \right).
\end{equation}
According to Fig.~\ref{fig:prob_cylinder_meanSquaredDisp_AiexVariations} this is valid for $A_{\mathrm{ex}}$ changes up to $\pm 10$\,\%. Typically, one assumes a distribution of the Curie temperature of 3\,\%\,$T_{\mathrm{C}}$. Hence, one can use Eq.~\ref{eq:scale_prob} to directly consider $T_{\mathrm{C}}$ distributions without the need to recalculate the switching probability for each variation of the exchange constant. 
\section{Conclusion}
\label{sec:conclusion}
To conclude, we presented an extensive study on how the material functions $\widetilde{\chi}_{\parallel}(T)$, $\widetilde{\chi}_{\perp}(T)$ and $m_{\mathrm{e}}(T)$, which are required to correctly integrate the Landau-Lifshitz-Bloch (LLB) equation (Eq.~\ref{eq:LLB}), depend on the size and the exchange constant of typical recording grains. The material functions for each system were extracted from atomistic Landau-Lifshitz-Gilbert (LLG) simulations. Further, we defined a reference particle and analyzed how scaling or shifting of its material curves, according to the changed Curie temperature, coincide with the separately computed ones. Additionally, we simulated a typical write process during heat-assisted recording (HAMR) and compared the resulting switching probabilities, based on the different input functions. 

We found that in the case of particle size variations the scaling and shifting approaches preform equally, within the investigated temperature range. The scaling and shifting approach well reproduce the correct $\widetilde{\chi}_{\parallel}(T)$, $\widetilde{\chi}_{\perp}(T)$ and $m_{\mathrm{e}}(T)$ curves as well as the correct switching probabilities for volume changes of up to $\pm 40$\,\%. The attempt to directly scale (or shift) the switching probability curve of the reference system (instead of recalculating them with scaled material functions) to the new Curie temperature yielded good results for volume changes of up to $\pm 10$\,\%.

For the variation of the exchange constant the scaling approach performed better than the shifting approach. The error was negligible for differences in the exchange constant of up to $\pm 10$\,\%, which corresponds to a $T_{\mathrm{C}}$ variation of more than $\pm 50$\,K. Direct scaling of the switching probabilities turned out to have similar errors. Against the background that typically a 3\,\% $T_{\mathrm{C}}$ distribution must be considered in HAMR simulations, this finding is important to significantly reduce computation time of bit-error rates whilst maintaining accuracy. Our results suggest the conclusion that switching probabilities does not need to be recalculated in HAMR studies if one considers $T_{\mathrm{C}}$ distribution. A simple scaling is sufficient.

\section{Acknowledgements}
The authors would like to thank the Vienna Science and Technology Fund (WWTF) under grant No. MA14-044, the Advanced Storage Technology Consortium (ASTC), and the Austrian Science Fund (FWF) under Grant Nos. F4112 SFB ViCoM and I2214-N20 for financial support. The support from the CD-laboratory AMSEN (financed by the Austrian Federal Ministry of Economy, Family and Youth, the National Foundation for Research, Technology and Development) was acknowledged. The computational results presented have been achieved using the Vienna Scientific Cluster (VSC).


\begin{thebibliography}{16}%
\makeatletter
\providecommand \@ifxundefined [1]{%
 \@ifx{#1\undefined}
}%
\providecommand \@ifnum [1]{%
 \ifnum #1\expandafter \@firstoftwo
 \else \expandafter \@secondoftwo
 \fi
}%
\providecommand \@ifx [1]{%
 \ifx #1\expandafter \@firstoftwo
 \else \expandafter \@secondoftwo
 \fi
}%
\providecommand \natexlab [1]{#1}%
\providecommand \enquote  [1]{``#1''}%
\providecommand \bibnamefont  [1]{#1}%
\providecommand \bibfnamefont [1]{#1}%
\providecommand \citenamefont [1]{#1}%
\providecommand \href@noop [0]{\@secondoftwo}%
\providecommand \href [0]{\begingroup \@sanitize@url \@href}%
\providecommand \@href[1]{\@@startlink{#1}\@@href}%
\providecommand \@@href[1]{\endgroup#1\@@endlink}%
\providecommand \@sanitize@url [0]{\catcode `\\12\catcode `\$12\catcode
  `\&12\catcode `\#12\catcode `\^12\catcode `\_12\catcode `\%12\relax}%
\providecommand \@@startlink[1]{}%
\providecommand \@@endlink[0]{}%
\providecommand \url  [0]{\begingroup\@sanitize@url \@url }%
\providecommand \@url [1]{\endgroup\@href {#1}{\urlprefix }}%
\providecommand \urlprefix  [0]{URL }%
\providecommand \Eprint [0]{\href }%
\providecommand \doibase [0]{http://dx.doi.org/}%
\providecommand \selectlanguage [0]{\@gobble}%
\providecommand \bibinfo  [0]{\@secondoftwo}%
\providecommand \bibfield  [0]{\@secondoftwo}%
\providecommand \translation [1]{[#1]}%
\providecommand \BibitemOpen [0]{}%
\providecommand \bibitemStop [0]{}%
\providecommand \bibitemNoStop [0]{.\EOS\space}%
\providecommand \EOS [0]{\spacefactor3000\relax}%
\providecommand \BibitemShut  [1]{\csname bibitem#1\endcsname}%
\let\auto@bib@innerbib\@empty
\bibitem [{\citenamefont {Garanin}(1997)}]{garanin_fokker-planck_1997}%
  \BibitemOpen
  \bibfield  {author} {\bibinfo {author} {\bibfnamefont {D.~A.}\ \bibnamefont
  {Garanin}},\ }\href {\doibase 10.1103/PhysRevB.55.3050} {\bibfield  {journal}
  {\bibinfo  {journal} {Phys. Rev. B}\ }\textbf {\bibinfo {volume} {55}},\
  \bibinfo {pages} {3050} (\bibinfo {year} {1997})}\BibitemShut {NoStop}%
\bibitem [{\citenamefont {Garanin}\ and\ \citenamefont
  {Chubykalo-Fesenko}(2004)}]{garanin_thermal_2004}%
  \BibitemOpen
  \bibfield  {author} {\bibinfo {author} {\bibfnamefont {D.~A.}\ \bibnamefont
  {Garanin}}\ and\ \bibinfo {author} {\bibfnamefont {O.}~\bibnamefont
  {Chubykalo-Fesenko}},\ }\href {\doibase 10.1103/PhysRevB.70.212409}
  {\bibfield  {journal} {\bibinfo  {journal} {Phys. Rev. B}\ }\textbf {\bibinfo
  {volume} {70}},\ \bibinfo {pages} {212409} (\bibinfo {year}
  {2004})}\BibitemShut {NoStop}%
\bibitem [{\citenamefont {Evans}\ \emph {et~al.}(2012)\citenamefont {Evans},
  \citenamefont {Hinzke}, \citenamefont {Atxitia}, \citenamefont {Nowak},
  \citenamefont {Chantrell},\ and\ \citenamefont
  {Chubykalo-Fesenko}}]{evans_stochastic_2012}%
  \BibitemOpen
  \bibfield  {author} {\bibinfo {author} {\bibfnamefont {R.~F.~L.}\
  \bibnamefont {Evans}}, \bibinfo {author} {\bibfnamefont {D.}~\bibnamefont
  {Hinzke}}, \bibinfo {author} {\bibfnamefont {U.}~\bibnamefont {Atxitia}},
  \bibinfo {author} {\bibfnamefont {U.}~\bibnamefont {Nowak}}, \bibinfo
  {author} {\bibfnamefont {R.~W.}\ \bibnamefont {Chantrell}}, \ and\ \bibinfo
  {author} {\bibfnamefont {O.}~\bibnamefont {Chubykalo-Fesenko}},\ }\href
  {\doibase 10.1103/PhysRevB.85.014433} {\bibfield  {journal} {\bibinfo
  {journal} {Phys. Rev. B}\ }\textbf {\bibinfo {volume} {85}},\ \bibinfo
  {pages} {014433} (\bibinfo {year} {2012})}\BibitemShut {NoStop}%
\bibitem [{\citenamefont {Chubykalo-Fesenko}\ \emph {et~al.}(2006)\citenamefont
  {Chubykalo-Fesenko}, \citenamefont {Nowak}, \citenamefont {Chantrell},\ and\
  \citenamefont {Garanin}}]{chubykalo-fesenko_dynamic_2006}%
  \BibitemOpen
  \bibfield  {author} {\bibinfo {author} {\bibfnamefont {O.}~\bibnamefont
  {Chubykalo-Fesenko}}, \bibinfo {author} {\bibfnamefont {U.}~\bibnamefont
  {Nowak}}, \bibinfo {author} {\bibfnamefont {R.~W.}\ \bibnamefont
  {Chantrell}}, \ and\ \bibinfo {author} {\bibfnamefont {D.}~\bibnamefont
  {Garanin}},\ }\href {\doibase 10.1103/PhysRevB.74.094436} {\bibfield
  {journal} {\bibinfo  {journal} {Phys. Rev. B}\ }\textbf {\bibinfo {volume}
  {74}},\ \bibinfo {pages} {094436} (\bibinfo {year} {2006})}\BibitemShut
  {NoStop}%
\bibitem [{\citenamefont {Bunce}\ \emph {et~al.}(2010)\citenamefont {Bunce},
  \citenamefont {Wu}, \citenamefont {Ju}, \citenamefont {Lu}, \citenamefont
  {Hinzke}, \citenamefont {Kazantseva}, \citenamefont {Nowak},\ and\
  \citenamefont {Chantrell}}]{bunce_laser-induced_2010}%
  \BibitemOpen
  \bibfield  {author} {\bibinfo {author} {\bibfnamefont {C.}~\bibnamefont
  {Bunce}}, \bibinfo {author} {\bibfnamefont {J.}~\bibnamefont {Wu}}, \bibinfo
  {author} {\bibfnamefont {G.}~\bibnamefont {Ju}}, \bibinfo {author}
  {\bibfnamefont {B.}~\bibnamefont {Lu}}, \bibinfo {author} {\bibfnamefont
  {D.}~\bibnamefont {Hinzke}}, \bibinfo {author} {\bibfnamefont
  {N.}~\bibnamefont {Kazantseva}}, \bibinfo {author} {\bibfnamefont
  {U.}~\bibnamefont {Nowak}}, \ and\ \bibinfo {author} {\bibfnamefont {R.~W.}\
  \bibnamefont {Chantrell}},\ }\href {\doibase 10.1103/PhysRevB.81.174428}
  {\bibfield  {journal} {\bibinfo  {journal} {Phys. Rev. B}\ }\textbf {\bibinfo
  {volume} {81}},\ \bibinfo {pages} {174428} (\bibinfo {year}
  {2010})}\BibitemShut {NoStop}%
\bibitem [{\citenamefont {Vogler}\ \emph {et~al.}(2014)\citenamefont {Vogler},
  \citenamefont {Abert}, \citenamefont {Bruckner},\ and\ \citenamefont
  {Suess}}]{volger_llb}%
  \BibitemOpen
  \bibfield  {author} {\bibinfo {author} {\bibfnamefont {C.}~\bibnamefont
  {Vogler}}, \bibinfo {author} {\bibfnamefont {C.}~\bibnamefont {Abert}},
  \bibinfo {author} {\bibfnamefont {F.}~\bibnamefont {Bruckner}}, \ and\
  \bibinfo {author} {\bibfnamefont {D.}~\bibnamefont {Suess}},\ }\href
  {\doibase 10.1103/PhysRevB.90.214431} {\bibfield  {journal} {\bibinfo
  {journal} {Phys. Rev. B}\ }\textbf {\bibinfo {volume} {90}},\ \bibinfo
  {pages} {214431} (\bibinfo {year} {2014})}\BibitemShut {NoStop}%
\bibitem [{\citenamefont {Atxitia}\ \emph {et~al.}(2007)\citenamefont
  {Atxitia}, \citenamefont {Chubykalo-Fesenko}, \citenamefont {Kazantseva},
  \citenamefont {Hinzke}, \citenamefont {Nowak},\ and\ \citenamefont
  {Chantrell}}]{atxitia_micromagnetic_2007}%
  \BibitemOpen
  \bibfield  {author} {\bibinfo {author} {\bibfnamefont {U.}~\bibnamefont
  {Atxitia}}, \bibinfo {author} {\bibfnamefont {O.}~\bibnamefont
  {Chubykalo-Fesenko}}, \bibinfo {author} {\bibfnamefont {N.}~\bibnamefont
  {Kazantseva}}, \bibinfo {author} {\bibfnamefont {D.}~\bibnamefont {Hinzke}},
  \bibinfo {author} {\bibfnamefont {U.}~\bibnamefont {Nowak}}, \ and\ \bibinfo
  {author} {\bibfnamefont {R.~W.}\ \bibnamefont {Chantrell}},\ }\href {\doibase
  10.1063/1.2822807} {\bibfield  {journal} {\bibinfo  {journal} {Appl. Phys.
  Lett.}\ }\textbf {\bibinfo {volume} {91}},\ \bibinfo {pages} {232507}
  (\bibinfo {year} {2007})}\BibitemShut {NoStop}%
\bibitem [{\citenamefont {Kazantseva}\ \emph {et~al.}(2008)\citenamefont
  {Kazantseva}, \citenamefont {Hinzke}, \citenamefont {Nowak}, \citenamefont
  {Chantrell}, \citenamefont {Atxitia},\ and\ \citenamefont
  {Chubykalo-Fesenko}}]{kazantseva_towards_2008}%
  \BibitemOpen
  \bibfield  {author} {\bibinfo {author} {\bibfnamefont {N.}~\bibnamefont
  {Kazantseva}}, \bibinfo {author} {\bibfnamefont {D.}~\bibnamefont {Hinzke}},
  \bibinfo {author} {\bibfnamefont {U.}~\bibnamefont {Nowak}}, \bibinfo
  {author} {\bibfnamefont {R.~W.}\ \bibnamefont {Chantrell}}, \bibinfo {author}
  {\bibfnamefont {U.}~\bibnamefont {Atxitia}}, \ and\ \bibinfo {author}
  {\bibfnamefont {O.}~\bibnamefont {Chubykalo-Fesenko}},\ }\href {\doibase
  10.1103/PhysRevB.77.184428} {\bibfield  {journal} {\bibinfo  {journal} {Phys.
  Rev. B}\ }\textbf {\bibinfo {volume} {77}},\ \bibinfo {pages} {184428}
  (\bibinfo {year} {2008})}\BibitemShut {NoStop}%
\bibitem [{\citenamefont {Schieback}\ \emph {et~al.}(2009)\citenamefont
  {Schieback}, \citenamefont {Hinzke}, \citenamefont {Kl{\"a}ui}, \citenamefont
  {Nowak},\ and\ \citenamefont {Nielaba}}]{schieback_temperature_2009}%
  \BibitemOpen
  \bibfield  {author} {\bibinfo {author} {\bibfnamefont {C.}~\bibnamefont
  {Schieback}}, \bibinfo {author} {\bibfnamefont {D.}~\bibnamefont {Hinzke}},
  \bibinfo {author} {\bibfnamefont {M.}~\bibnamefont {Kl{\"a}ui}}, \bibinfo
  {author} {\bibfnamefont {U.}~\bibnamefont {Nowak}}, \ and\ \bibinfo {author}
  {\bibfnamefont {P.}~\bibnamefont {Nielaba}},\ }\href {\doibase
  10.1103/PhysRevB.80.214403} {\bibfield  {journal} {\bibinfo  {journal} {Phys.
  Rev. B}\ }\textbf {\bibinfo {volume} {80}},\ \bibinfo {pages} {214403}
  (\bibinfo {year} {2009})}\BibitemShut {NoStop}%
\bibitem [{\citenamefont {McDaniel}(2012)}]{mcdaniel_application_2012}%
  \BibitemOpen
  \bibfield  {author} {\bibinfo {author} {\bibfnamefont {T.~W.}\ \bibnamefont
  {McDaniel}},\ }\href {\doibase 10.1063/1.4733311} {\bibfield  {journal}
  {\bibinfo  {journal} {J. Appl. Phys.}\ }\textbf {\bibinfo {volume} {112}},\
  \bibinfo {pages} {013914} (\bibinfo {year} {2012})}\BibitemShut {NoStop}%
\bibitem [{\citenamefont {Greaves}\ \emph {et~al.}(2012)\citenamefont
  {Greaves}, \citenamefont {Kanai},\ and\ \citenamefont
  {Muraoka}}]{greaves_magnetization_2012}%
  \BibitemOpen
  \bibfield  {author} {\bibinfo {author} {\bibfnamefont {S.}~\bibnamefont
  {Greaves}}, \bibinfo {author} {\bibfnamefont {Y.}~\bibnamefont {Kanai}}, \
  and\ \bibinfo {author} {\bibfnamefont {H.}~\bibnamefont {Muraoka}},\ }\href
  {\doibase 10.1109/TMAG.2012.2187776} {\bibfield  {journal} {\bibinfo
  {journal} {{IEEE} Trans. Magn.}\ }\textbf {\bibinfo {volume} {48}},\ \bibinfo
  {pages} {1794} (\bibinfo {year} {2012})}\BibitemShut {NoStop}%
\bibitem [{\citenamefont {Mendil}\ \emph {et~al.}()\citenamefont {Mendil},
  \citenamefont {Nieves}, \citenamefont {Chubykalo-Fesenko}, \citenamefont
  {Walowski}, \citenamefont {Santos}, \citenamefont {Pisana},\ and\
  \citenamefont {M{\"u}nzenberg}}]{mendil_resolving_2014}%
  \BibitemOpen
  \bibfield  {author} {\bibinfo {author} {\bibfnamefont {J.}~\bibnamefont
  {Mendil}}, \bibinfo {author} {\bibfnamefont {P.}~\bibnamefont {Nieves}},
  \bibinfo {author} {\bibfnamefont {O.}~\bibnamefont {Chubykalo-Fesenko}},
  \bibinfo {author} {\bibfnamefont {J.}~\bibnamefont {Walowski}}, \bibinfo
  {author} {\bibfnamefont {T.}~\bibnamefont {Santos}}, \bibinfo {author}
  {\bibfnamefont {S.}~\bibnamefont {Pisana}}, \ and\ \bibinfo {author}
  {\bibfnamefont {M.}~\bibnamefont {M{\"u}nzenberg}},\ }\href {\doibase
  10.1038/srep03980} {\bibfield  {journal} {\bibinfo  {journal} {Sci. Rep.}\
  }\textbf {\bibinfo {volume} {4}},\ 10.1038/srep03980}\BibitemShut {NoStop}%
\bibitem [{\citenamefont {Evans}\ \emph {et~al.}(2014)\citenamefont {Evans},
  \citenamefont {Fan}, \citenamefont {Chureemart}, \citenamefont {Ostler},
  \citenamefont {Ellis},\ and\ \citenamefont
  {Chantrell}}]{evans_atomistic_2014}%
  \BibitemOpen
  \bibfield  {author} {\bibinfo {author} {\bibfnamefont {R.~F.~L.}\
  \bibnamefont {Evans}}, \bibinfo {author} {\bibfnamefont {W.~J.}\ \bibnamefont
  {Fan}}, \bibinfo {author} {\bibfnamefont {P.}~\bibnamefont {Chureemart}},
  \bibinfo {author} {\bibfnamefont {T.~A.}\ \bibnamefont {Ostler}}, \bibinfo
  {author} {\bibfnamefont {M.~O.~A.}\ \bibnamefont {Ellis}}, \ and\ \bibinfo
  {author} {\bibfnamefont {R.~W.}\ \bibnamefont {Chantrell}},\ }\href {\doibase
  10.1088/0953-8984/26/10/103202} {\bibfield  {journal} {\bibinfo  {journal}
  {J. Phys.: Condens. Matter}\ }\textbf {\bibinfo {volume} {26}},\ \bibinfo
  {pages} {103202} (\bibinfo {year} {2014})}\BibitemShut {NoStop}%
\bibitem [{\citenamefont {Ellis}\ and\ \citenamefont
  {Chantrell}()}]{ellis_switching_2015}%
  \BibitemOpen
  \bibfield  {author} {\bibinfo {author} {\bibfnamefont {M.~O.~A.}\
  \bibnamefont {Ellis}}\ and\ \bibinfo {author} {\bibfnamefont {R.~W.}\
  \bibnamefont {Chantrell}},\ }\href {\doibase 10.1063/1.4919051} {\bibfield
  {journal} {\bibinfo  {journal} {Appl. Phys. Lett.}\ }\textbf {\bibinfo
  {volume} {106}},\ \bibinfo {pages} {162407}}\BibitemShut {NoStop}%
\bibitem [{\citenamefont {Hovorka}\ \emph {et~al.}(2012)\citenamefont
  {Hovorka}, \citenamefont {Devos}, \citenamefont {Coopman}, \citenamefont
  {Fan}, \citenamefont {Aas}, \citenamefont {Evans}, \citenamefont {Chen},
  \citenamefont {Ju},\ and\ \citenamefont {Chantrell}}]{hovorka_curie_2012}%
  \BibitemOpen
  \bibfield  {author} {\bibinfo {author} {\bibfnamefont {O.}~\bibnamefont
  {Hovorka}}, \bibinfo {author} {\bibfnamefont {S.}~\bibnamefont {Devos}},
  \bibinfo {author} {\bibfnamefont {Q.}~\bibnamefont {Coopman}}, \bibinfo
  {author} {\bibfnamefont {W.~J.}\ \bibnamefont {Fan}}, \bibinfo {author}
  {\bibfnamefont {C.~J.}\ \bibnamefont {Aas}}, \bibinfo {author} {\bibfnamefont
  {R.~F.~L.}\ \bibnamefont {Evans}}, \bibinfo {author} {\bibfnamefont
  {X.}~\bibnamefont {Chen}}, \bibinfo {author} {\bibfnamefont {G.}~\bibnamefont
  {Ju}}, \ and\ \bibinfo {author} {\bibfnamefont {R.~W.}\ \bibnamefont
  {Chantrell}},\ }\href {\doibase 10.1063/1.4740075} {\bibfield  {journal}
  {\bibinfo  {journal} {Appl. Phys. Lett.}\ }\textbf {\bibinfo {volume}
  {101}},\ \bibinfo {pages} {052406} (\bibinfo {year} {2012})}\BibitemShut
  {NoStop}%
\bibitem [{\citenamefont {Binder}()}]{binder_applications_1997}%
  \BibitemOpen
  \bibfield  {author} {\bibinfo {author} {\bibfnamefont {K.}~\bibnamefont
  {Binder}},\ }\href {\doibase 10.1088/0034-4885/60/5/001} {\bibfield
  {journal} {\bibinfo  {journal} {Rep. Prog. Phys.}\ }\textbf {\bibinfo
  {volume} {60}},\ \bibinfo {pages} {487}}\BibitemShut {NoStop}%
\end{thebibliography}
\end{document}